\documentclass[12pt,thmsa]{article}
\usepackage{sw20lart}

\input tcilatex
\QQQ{Language}{
American English
}

\begin{document}

\author{I. Radinschi \and ''Gh. Asachi'' Technical University, Department of
Physics, \and B-dul Dimitrie Mangeron, No. 67, Iasi, Romania, 6600 \and %
jessica@etc.tuiasi.ro}
\title{The Energy of a Dyonic Dilaton Black Hole }
\date{The Date }
\maketitle

\begin{abstract}
We calculate the energy distribution of a dyonic dilaton black hole by using
the Tolman's energy-momentum complex. All the calculations are performed in
quasi-Cartesian coordinates. The energy distribution of the dyonic dilaton
black hole depends on the mass $M$, electric charge $Q_e$, magnetic charge $%
Q_m$ and asymptotic value of the dilaton $\Phi _0$. We get the same result
as obtained by Y-Ching Yang, Ching-Tzung Yeh, Rue-Ron Hsu and Chin-Rong Lee
by using the Einstein's prescription.


\end{abstract}

\section{INTRODUCTION}

The energy-momentum localization has been a problematic issue since the
outset of the theory of relativity. A large number of definitions of the
gravitational energy have been given since now. Some of them are coordinate
independent and other are coordinate-dependent. An adequate
coordinate-independent prescription for energy-momentum localization for all
the type of space-times has not given yet in General Relativity.

We remark that it is possible to evaluate the energy and momentum
distribution by using various energy-momentum complexes. The physical
interpretation of these nontensorial energy-momentum complexes have been
questioned by a number of physicists, including Weyl, Pauli and Eddington.
There prevails suspicion that different energy-momentum complexes could give
different energy distributions in a given space-time. Virbhadra and his
collaborators have considered many space-times and have shown that several
energy-momentum complexes give the same and acceptable result for a given
space-time.

Many authors obtained dilaton black hole solutions and studied theirs
properties [1]-[4]. Garfinkle, Horowitz and Strominger (GHS) [5] obtained a
form of static spherically symmetric charged dilaton black hole solutions
which exhibit several different properties compared to the
Reissner-Nordstr\"{o}m (RN) black holes. In their theory the gravity is
coupled to the electromagnetic and dilaton fields and can be described by
the four-dimensional effective string action.

Chamorro and Virbhadra [6] obtained in the Einstein's prescription the
energy of a charged dilaton black hole based on the GHS [5] solutions. They
found that the energy distribution which has the expression $E(r)=M-\frac{Q^2%
}{2r}(1-\beta ^2)$, depends on the mass $M$, electric charge $Q_e$ and the
coupling parameter $\beta $ between the dilaton and the Maxwell fields. For
the value $\beta =0$ they obtained the energy distribution in the
Reissner-Nordstr\"{o}m (RN) field. Also, only for $\beta =1$ the energy is
confined to its interior, and for all other values of $\beta $ the energy is
shared by the interior and exterior of the black holes. The total energy of
the charged dilaton black hole is independent of $\beta $ and is given by
the mass parameter of the black hole. With increasing the radial distance, $%
E(r)$ increases for $\beta =0$ (RN metric) as well for $\beta <1$, decreases
for $\beta >1$, and remains constant for $\beta =1$.

S. S. Xulu [7] get the same energy distribution as Chamorro and Virbhadra
[6] by using the Tolman's prescription. The energy distribution that is
given by $E(r)=M-\frac{Q^2}{2r}(1-\beta ^2)$ can be interpreted as the
''effective gravitational mass'' that a neutral test particle ''feels'' in
the GHS space-time. Also, the ''effective gravitational mass'' becomes
negative at radial distances less than $\frac{Q^2}{2M}(1-\beta ^2)$.

Virbhadra and Parikh [8] investigated, in the Einstein's prescription [9],
the energy of a static spherically symmetric charged dilaton black hole and
found that the entire energy is confined to its interior with no energy
shared by the exterior of the black hole. This result is similar to the case
of the Schwarzschild black hole and unlike the RN black hole.

Cheng, Lin and Hsu (CLH) [10] using the standard spherical coordinate system
which is more suitable for describing the structure of the charged dilaton
black hole, obtained the more general solutions which are the dyonic dilaton
black hole solutions. The GHS solutions can be obtained from the CLH
solutions as special cases when electric or magnetic charges are switched
off.

I-Ching Yang, Ching-Tzung Yeh, Rue-Ron Hsu and Chin-Rong Lee [11] employing
the Einstein's energy-momentum complex obtained that the energy distribution
of a dyonic dilaton black hole depends on the mass $M$, electric charge $Q_e$%
, magnetic charge $Q_m$ and asymptotic value of the dilaton $\Phi _0$.

In this paper we compute the energy distribution of a dyonic dilaton black
hole by using the Tolman's prescription [12] . We obtain the same result as
obtained by I. Ching-Yang, Ching-Tzung Yeh, Rue-Ron Hsu and Chin-Rong Lee
[11]. We also make a discussion of the results. We use the geometrized units 
$(G=1,c=1)$ and follow the convention that the Latin indices run from $0$ to 
$3$.

\section{THE\ ENERGY\ DISTRIBUTION IN\ THE\ TOLMAN'S\ PRESCRIPTION}

The static, spherical symmetric dyonic dilaton black hole solutions are in
terms of the standard spherical coordinate [10]

\begin{equation}
ds^2=\Delta ^2dt^2-\frac{\sigma ^2}{\Delta ^2}dr^2-r^2d\theta ^2-r^2\sin
^2\theta d\varphi ^2
\end{equation}

where

\[
\sigma ^2=\frac{r^2}{r^2+\lambda ^2}, 
\]

\[
\Delta ^2=1-\frac{2M}{r^2}\sqrt{r^2+\lambda ^2}+\frac \beta {r^2}, 
\]

\begin{equation}
\lambda =\frac 1{2M}(Q_e^2e^{2\Phi _0}-Q_m^2e^{-2\Phi _0}),
\end{equation}

\[
\beta =Q_e^2e^{2\Phi _0}+Q_m^2e^{-2\Phi _0}, 
\]

\[
e^{2\Phi }=e^{-2\Phi _0}(1-\frac{2\lambda }{\sqrt{r^2+\lambda ^2}+\lambda }%
). 
\]

The only non-zero components of the electromagnetic field tensor are

\begin{equation}
F_{01}=\frac{Q_e}{r^2}e^{2\Phi }
\end{equation}

and, respectively

\begin{equation}
F_{23}=\frac{Q_m}{r^2}.
\end{equation}

The properties of the dyonic dilaton black holes are characterized by the
mass $M$, electric charge $Q_e$, magnetic charge $Q_m$ and asymptotic value
of the dilaton $\Phi _0$. Their structures are similar to that of the RN
[10] black holes.

The Tolman's energy-momentum complex [12] is given by

\begin{equation}
\Upsilon _i^{\;\;k}={\frac 1{8\pi }}U_i^{\;kl},_l,
\end{equation}

where $\Upsilon _0^{\;\;0}$ and $\Upsilon _\alpha ^{\;\;0}$ are the energy
and momentum components.

We have

\begin{equation}
U_i^{\;kl}=\sqrt{-g}(-g^{pk}V_{ip}^{\;~l}+\frac
12g_i^kg^{pm}V_{pm}^{\;~\;\;l}),
\end{equation}

with

\begin{equation}
V_{jk}^{\;\;i}=-\Gamma _{jk}^i+{\frac 12}g_j^i\Gamma _{mk}^m+{\frac 12}%
g_k^i\Gamma _{mj}^m.
\end{equation}

The energy-momentum complex $\Upsilon _i^{\;\;k}$ also satisfies the local
conservation laws

\begin{equation}
\frac{\partial \Upsilon _i^{\;\;k}}{\partial x^k}=0.
\end{equation}

The Tolman's energy-momentum complex gives the correct result if the
calculations are carried out in quasi-Cartesian coordinates.

We transform the line element (1) to quasi-Cartesian coordinates $t,x,y,z$
according to

\begin{equation}
\begin{tabular}{c}
$x=r\sin \theta \cos \varphi ,$ \\ 
$y=r\sin \theta \sin \varphi ,$ \\ 
$z=r\cos \theta $%
\end{tabular}
\end{equation}

and

\begin{equation}
r=(x^2+y^2+z^2)^{\frac 12}.
\end{equation}

The line element (1) becomes

\begin{equation}
ds^2=\Delta ^2dt^2-(dx^2+dy^2+dz^2)-\frac{\sigma ^2/\Delta ^2-1}{r^2}%
(xdx+ydy+zdz)^2.
\end{equation}

The only required components of $U_i^{\;kl}$ in the calculation of the
energy are the following

\[
U_0^{\;01}=\frac{x\Pi }{r^2}, 
\]

\begin{equation}
U_0^{\;02}=\frac{y\Pi }{r^2},
\end{equation}

\[
U_0^{\;03}=\frac{z\Pi }{r^2}. 
\]

In the relations (12) we denote by $\Pi $

\begin{equation}
\Pi =\sigma (1-\frac{\Delta ^2}{\sigma ^2}).
\end{equation}

The components of the pseudotensor $U_i^{\;kl}$ are calculated with the
program Maple GR Tensor II Release 1.50.

The energy and momentum in the Tolman's prescription are given by

\begin{equation}
P_i=\iiint \Upsilon _i^{\;0}dx^1dx^2dx^3.
\end{equation}

Using the Gauss's theorem we obtain

\begin{equation}
P_i=\frac 1{8\pi }\iint U_i^{\;0\alpha }n_\alpha dS,
\end{equation}

where $n_\alpha =(x/r,y/r,z/r)$ are the components of a normal vector over
an infinitesimal surface element $dS=r^2\sin \theta d\theta d\varphi $.

Using (2), (12), (14) and applying the Gauss's theorem we evaluate the
integral over the surface of a sphere with radius $r$

\begin{equation}
E(r)=\frac 1{8\pi }\oint \frac \sigma r(1-\frac{\Delta ^2}{\sigma ^2}%
)r^2\sin \theta d\theta d\varphi .
\end{equation}

We find that the energy within a sphere with radius $r$ is given

\begin{equation}
E(r)=M+\frac{M\lambda ^2}{r^2}-\frac 1{2\sqrt{r^2+\lambda ^2}}[\frac{\beta
\lambda ^2}{r^2}+\lambda ^2+\beta ].
\end{equation}

The energy distribution depends on the mass $M$, electric charge $Q_e$,
magnetic charge $Q_m$ and asymptotic value of the dilaton $\Phi _0$.

The energy is shared both by the interior and by the exterior of the black
hole.

\section{DISCUSSION}

The subject of the localization of energy continues to be an open one. Bondi
[13] sustained that a nonlocalizable form of energy is not admissible in
relativity. Other authors consider that because the energy-momentum
complexes are not tensorial objects and give results which are coordinate
dependent they are not adequate for describing the gravitational field.

The results obtained by Chamorro and Virbhadra [6], Xulu [7], Virbhadra and
Parikh [8] and I-Ching Yang, Ching-Tzung Yeh, Rue-Ron Hsu and Chin-Rong Lee
[11] support the idea that several energy-momentum complexes can give the
same result for a given space-time.

We obtain the same result as obtained by Y-Ching Yang, Ching-Tzung Yeh,
Rue-Ron Hsu and Chin-Rong Lee [11]. This is an encouraging result and, also,
it is one more proof that the Einstein's and Tolman's energy-momentum
complexes can give the same result for a static spherically symmetric
solution. The energy distribution depends on the mass $M$, electric charge $%
Q_e$, magnetic charge $Q_m$ and asymptotic value of the dilaton $\Phi _0$.
The energy is shared both by the interior and by the exterior of the black
hole.

We obtain the same expression for the energy distribution as in the case of
Schwarzschild black holes as $Q_e,Q_m$ and $\Phi _0$ vanish.

For $Q_e=0$ or $Q_m=0$ we find the case of the pure electric or pure
magnetic charged black hole and the energy distribution is always positive
except at the singular point $r=0$.

If the dilaton field was suppressed $\Phi _0=0$, $\lambda =0$ (or $Q_e=Q_m$)
the dilaton gravity will reduce to the Einstein-Maxwell theory and the
dilaton dyonic black hole solutions will become to be Reissner-Nordstr\"{o}m
solutions.

In the case of the ADM mass we obtain the same result as the result of
Virbhadra

\begin{equation}
M_{ADM}=E(r)_{r\rightarrow \infty }=M
\end{equation}

From (18) we deduce that the ADM mass does not depend on the coordinate
representation of the black hole.

Also, the concept of a black hole lends support to the idea that the
gravitational energy is localizable.

\end{document}